\documentclass{emulateapj}
\usepackage{graphicx}
\usepackage{natbib}

\newcommand{\msun}{M$_{\sun}$}
\newcommand{\rsun}{R$_{\sun}$}
\newcommand{\ldl}{$\lambda/{\Delta}{\lambda}$}
\newcommand{\teff}{T$_{eff}$}
\newcommand{\logg}{$\log{g}$}

\newcommand{\kms}{km~s$^{-1}$}

\newcommand{\teffresult}{2550$^{+80}_{-90}$}
\newcommand{\loggresult}{4.4$^{+0.5}_{-0.5}$}
\newcommand{\zresult}{$-$0.96$^{+0.19}_{-0.24}$}

\slugcomment{Submitted to AJ 18 September 2015\\ Accepted 24 December 2015}

\begin{document}

\title{Characterization of the Very Low Mass Secondary in the GJ~660.1AB System}
\author{Christian Aganze\altaffilmark{1,2}, 
Adam J.\ Burgasser\altaffilmark{2}, 
Jacqueline K.\ Faherty\altaffilmark{3,4,5}, 
Caleb Choban\altaffilmark{2}, 
Ivanna Escala\altaffilmark{2}, 
Mike A.\ Lopez\altaffilmark{2}, 
Yuhui Jin\altaffilmark{2}, 
Tomoki Tamiya\altaffilmark{2}, 
Melisa Tallis\altaffilmark{2}, 
and Willie Rockward\altaffilmark{1}}

\affil{\altaffilmark{1}Department of Physics and Dual-Degree Engineering, Morehouse College, 830 Westview Drive S.W, Atlanta, GA 30314, USA; christian.aganze@morehouse.edu}
\affil{\altaffilmark{2}Center for Astrophysics and Space Science, University of California San Diego, 9500 Gilman Dr, La Jolla, CA 92093, USA}
\affil{\altaffilmark{3}Department of Terrestrial Magnetism, Carnegie Institution of Washington, 5241 Broad Branch Road, N. W. Washington, DC 20015, USA}
\affil{\altaffilmark{4}Department of Astrophysics, American Museum of Natural History, Central Park West at 79th Street, New York, NY 10024, USA }
\affil{\altaffilmark{5}Hubble Postdoctoral Fellow}

\begin{abstract}
We present spectroscopic analysis of the low mass binary star system GJ~660.1AB, a pair of nearby M dwarfs for which we have obtained {separated} near-infrared spectra (0.9-2.5~$\micron$) with the SpeX spectrograph. The spectrum of GJ~660.1B is distinctly peculiar, with a triangular-shaped 1.7~$\micron$ peak that initially suggests it to be a low surface gravity, young brown dwarf. However, we rule out this hypothesis and determine instead that this companion is a mild subdwarf (d/sdM7) based on the subsolar metallicity of the primary, [Fe/H] = {$-$0.63$\pm$0.06}.
Comparison of the near-infrared spectrum of GJ~660.1B to two sets of spectral models yields conflicting results,
with a {common} effective temperature {\teff} = 2550--2650~K, but alternately  
low surface gravity ({\logg} = {\loggresult}) 
and very low metallicity ([M/H] = {\zresult}), or 
high surface gravity ({\logg} = 5.0--5.5)
and slightly subsolar metallicity ([M/H] =$-$0.20$^{+0.13}_{-0.19}$).
We conjecture that insufficient condensate opacity and excessive 
collision induced H$_2$ absorption in the models bias them toward low surface gravities
{and a metallicity inconsistent with the primary, and points toward improvements needed
in the spectral modeling of metal-poor, very-low mass dwarfs}. 
The {peculiar spectral characteristics of GJ~660.1B}
emphasize that care is needed when {interpreting} surface gravity features 
in the spectra of ultracool dwarfs.
\end{abstract}

\keywords{
stars: individual (\objectname{GJ~660.1A}, \objectname{GJ~660.1B}) --- 
stars: late-type ---
stars: low mass, brown dwarfs}

\section{Introduction}

Our understanding of brown dwarfs and very low mass (VLM; M $\lesssim$ 0.1~{\msun}) stars has advanced considerably since the first discovery of substellar objects in the 1990s \citep{1995Natur.377..129R,1995Natur.378..463N}. Subsequent searches in various wide-field, multi-band infrared imaging surveys have revealed thousands of low-temperature VLM {dwarfs, many in the immediate} vicinity of the Sun ($d$ $\lesssim$ 30~pc; \citealt{2005ARA&A..43..195K}).  Among these, VLM companions to nearby stars have been {particularly valuable}, as they share the distance, age and composition of their more massive, and more precisely characterized, primaries.  Many of these VLM ``benchmarks'' have been intensely studied, serving as empirical tests of spectral, evolutionary and formation models (e.g., \citealt{1988Natur.336..656B,1995Natur.378..463N,2000ApJ...531L..57B,2001AJ....121.3235K,2001AJ....122.1989W,2005A&A...438L..25C,2006MNRAS.373L..31M,2006ApJ...651.1166M,2007ApJ...656.1136S,2008ApJ...682.1256L,2010ApJ...720..252L,2009ApJ...699..168D,2010AJ....139..176F,2011AJ....141...71F,2011MNRAS.414.3590B,2012ApJ...753..142B,2012ApJ...756...69B,2013ApJ...774...55B,2013MNRAS.431.2745G}). 

\cite{2014ApJ...792..119D} have recently compiled a sample of VLM companions to nearby stars, many identified in the Pan-STARRS survey \citep{2002SPIE.4836..154K}. One system not included in this compilation is GJ~660.1AB, identified by \cite{2011ApJ...743..109S} as an M1 dwarf \citep{2003AJ....126.2048G} with a relatively faint ($\Delta{J}$ = 4) co-moving secondary at a separation of 6$\farcs$08$\pm$0$\farcs$02 (122$\pm$9~AU). \cite{2011ApJ...743..109S} estimated a spectral type of  M9$\pm$2  for the companion based on its relative brightness. The system has a parallactic distance of 20.0$\pm$1.4~pc, a high proper motion ($\vec\mu$ = [181$\pm$5, $-$694$\pm$3]~mas/yr; \citealt{2007A&A...474..653V}), and a radial velocity of  $-$33$\pm$10~{\kms} \citep{1996AJ....112.2799H}. As the Galactic velocity of the system, ($U, V, W$) = (0.5$\pm$2.1, $-$52$\pm$3, $-$60$\pm$4)~{\kms}, is inconsistent with any nearby young associations (e.g., \citealt{2004ARA&A..42..685Z,2007IAUS..237..442M,2008hsf2.book..757T})
and indicates old disk kinematics \citep{1992ApJS...82..351L}, Schneider et al.\ estimate the age of the system to be older than $\sim$2~Gyr. No spectral observations of  the companion have yet been reported.

In this article we {report separated}, near-infrared spectra for GJ~660.1A and B obtained with the NASA Infrared Telescope Facility (IRTF) SpeX spectrograph (\citealt{2003PASP..115..362R}). Section 2 summarizes the observations. Section 3 describes our analysis of the spectra, including  classification of the low-resolution data, characterization of spectral peculiarities, and analysis of moderate resolution spectra of the primary which yields the system's metallicity. Section 4 presents model fitting of the secondary spectrum and corresponding estimates of the effective temperature ({\teff}), surface gravity ({\logg}), metallicity ([M/H]) and other physical properties of this source. We discuss our findings in Section 5, focusing in particular on mismatches between the model and observed spectra and their possible origin.  For reference, Table~\ref{tab:properties} summarizes the properties of the GJ~660.1 components based on this and prior work.

\begin{deluxetable*}{lccc}
\tabletypesize{\small}
\tablecaption{Properties of the GJ~660.1AB System\label{tab:properties}}
\tablewidth{0pc}
\tablehead {
\colhead{Property} & \colhead{GJ~660.1A} & \colhead{GJ~660.1B} & \colhead{Ref} \\
}
\startdata
Spectral Type & M1 & d/sdM7 & 1,2 \\
Distance (pc) & \multicolumn{2}{c}{20.0$\pm$1.4}  &3 \\ 
2MASS $J$ &8.66$\pm$0.03 & 13.05$\pm$0.05& 4 \\ 
2MASS $H$ & 8.07$\pm$0.04 &12.57$\pm$0.02& 4 \\ 
2MASS $K_s$ & 7.94$\pm$0.02 &12.23$\pm$0.03 & 4 \\ 
Absolute 2MASS $J$ & 7.15$\pm$0.15 &11.54$\pm$0.15 & 3,4 \\ 
Absolute 2MASS $H$ & 6.56$\pm$0.15 & 11.06$\pm$0.15 & 3,4\\ 
Absolute 2MASS $K_s$ & 6.43$\pm$0.15 &10.72$\pm$0.15 & 3,4\\ 
Separation ($\arcsec$) &\multicolumn{2}{c}{6.08$\pm$0.02} & 2\\ 
Projected separation (AU) &\multicolumn{2}{c}{122$\pm$9} & 2 \\ 
{\teff} (K) & 3800 & {\teffresult}\tablenotemark{a} & 1,5 \\ 
{\logg} (cgs) & 4.74 & \nodata\tablenotemark{b} & 1,6 \\ 
$[M/H]$ & $-$0.47$\pm$0.07\tablenotemark{c} & {\zresult}\tablenotemark{a} & 1 \\ 
$[Fe/H]$ & $-$0.63$\pm$0.06\tablenotemark{c} & \nodata & 1 \\ 
%Position angle ($\degr$)& \multicolumn{c}{353.1$\pm$0.1} &2 \\ 
Radius ({\rsun})& 0.53$\pm$0.04  & 0.085$\pm$0.014\tablenotemark{a} & 1,6 \\ 
Mass ({\msun}) & 0.57$\pm$0.07 & 0.084--0.091 &  1,6  \\
%Age (Gyr)\tablenotemark{a} & $\gtrsim$ 2  & $1.250 ^{+7.956} _{-2.397}$  & 1,2\\
%Luminosity (L$_\sun$) & &$-2.817 ^{+0.873} _{-1.195}$ & 1\\
\enddata
\tablenotetext{a}{Based on  BT-Settl spectral model fit; the radius is based on the scale factor between models and data and includes uncertainty in distance; see Section 3.3.}
\tablenotetext{b}{{\logg} value from spectral model fits unreliable; see Section 4.}
\tablenotetext{c}{Based on analysis of SXD data; see Section~\ref{sec:metal-poor} and Table~\ref{tab:metallicity}.}
\tablerefs{(1) This paper; (2) \citet{2011ApJ...743..109S}; (3) \citet{2007A&A...474..653V}; (4) \citet{2003yCat.2246....0C}; (5) \citet{2014MNRAS.443.2561G}; (6) \citet{2015A&A...577A.132M}.}
\end{deluxetable*}

\section{Observations}

Observations of the GJ~660.1AB system with SpeX were made during the nights of 2011 March 9 and 2015 April 9 (UT). Conditions were clear on both nights, with 0$\farcs$8 and 0$\farcs$6 seeing at $K$-band, respectively. Our 2011 data were obtained with the SpeX prism mode and 0$\farcs$5 slit, aligned perpendicular to the binary axis (position angle = 315$\degr$), which provided 0.7--2.5~$\micron$ coverage in a single order for each component separately with resolution {\ldl} $\approx$ 120 and dispersion of 20--30 {\AA}~pixel$^{-1}$. GJ~660.1B was exposed for 6 images of 90~s each at an airmass of 1.157 while GJ~660.1A was exposed for 8 images for 1~s x 4 coadds each at an airmass of 1.139. We also observed the A0~V star calibrator HD~148968 (B=7.116, V=6.98) for flux calibration, and obtained flat field and HeNeAr arc lamps. 
In 2015, we observed the primary only with the SpeX cross dispersed mode and 0$\farcs$8$\times$15$\arcsec$ slit aligned at the parallactic angle, providing moderate resolution data ({\ldl} $\approx$ 750) with dispersion 3.6~{\AA}~pixel$^{-1}$ covering 0.8--2.4~$\micron$ in six orders. Four exposures of 90~s each were obtained at an airmass of  1.139, followed by the same A0~V calibrator star and lamp exposures.
Data reduction was performed using the SpeXtool package version 3.4 \citep{2003PASP..115..389V,2004PASP..116..362C}, which removes background, bad pixels and instrumental effects and optimally extracts the spectral data. The wavelength scale was calibrated to air wavelengths using the arc lamp spectrum, and slit and telluric transmission losses were corrected using the A0~V star observation. 

\begin{figure}
\epsscale{0.75}
\plotone{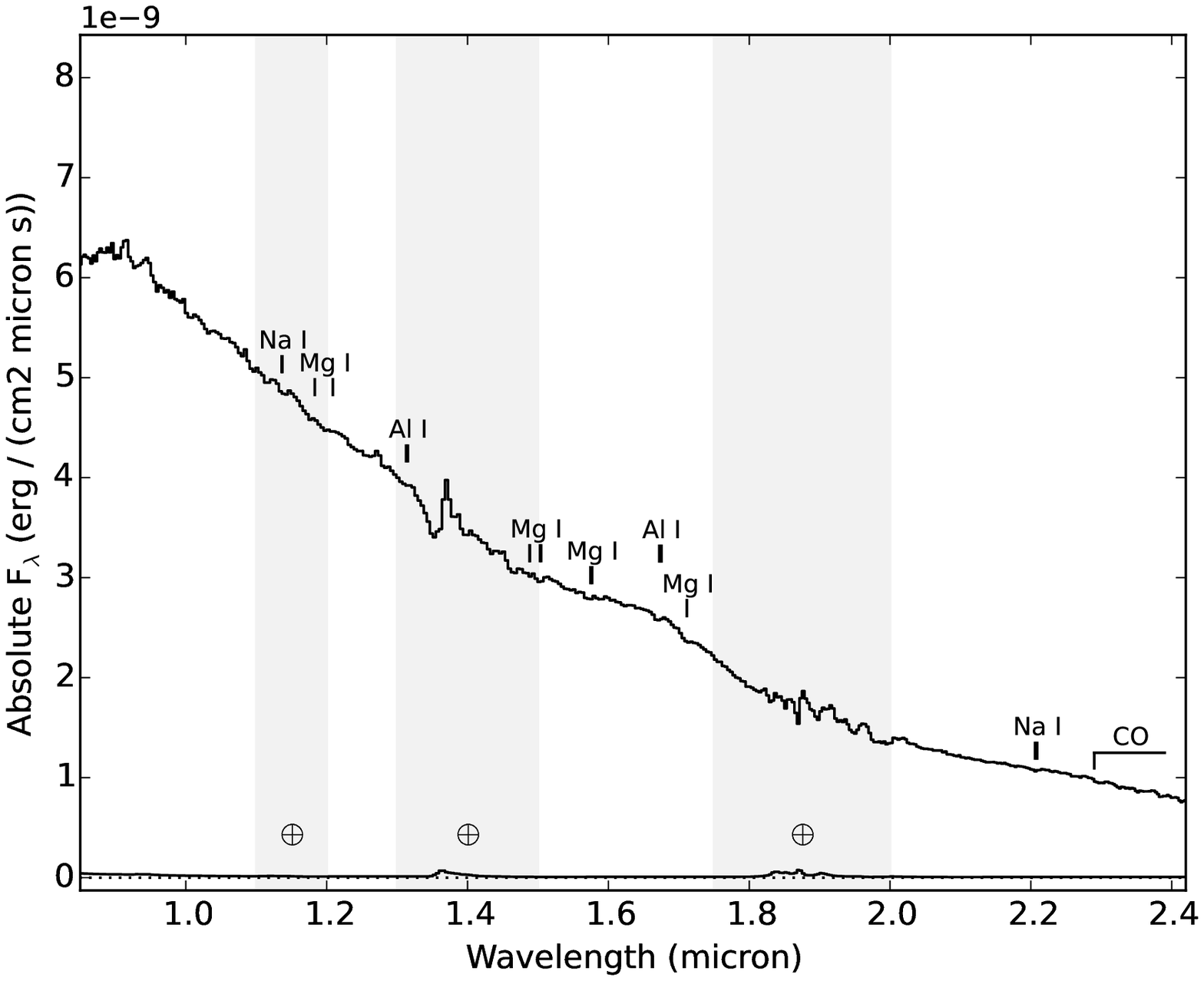}
\plotone{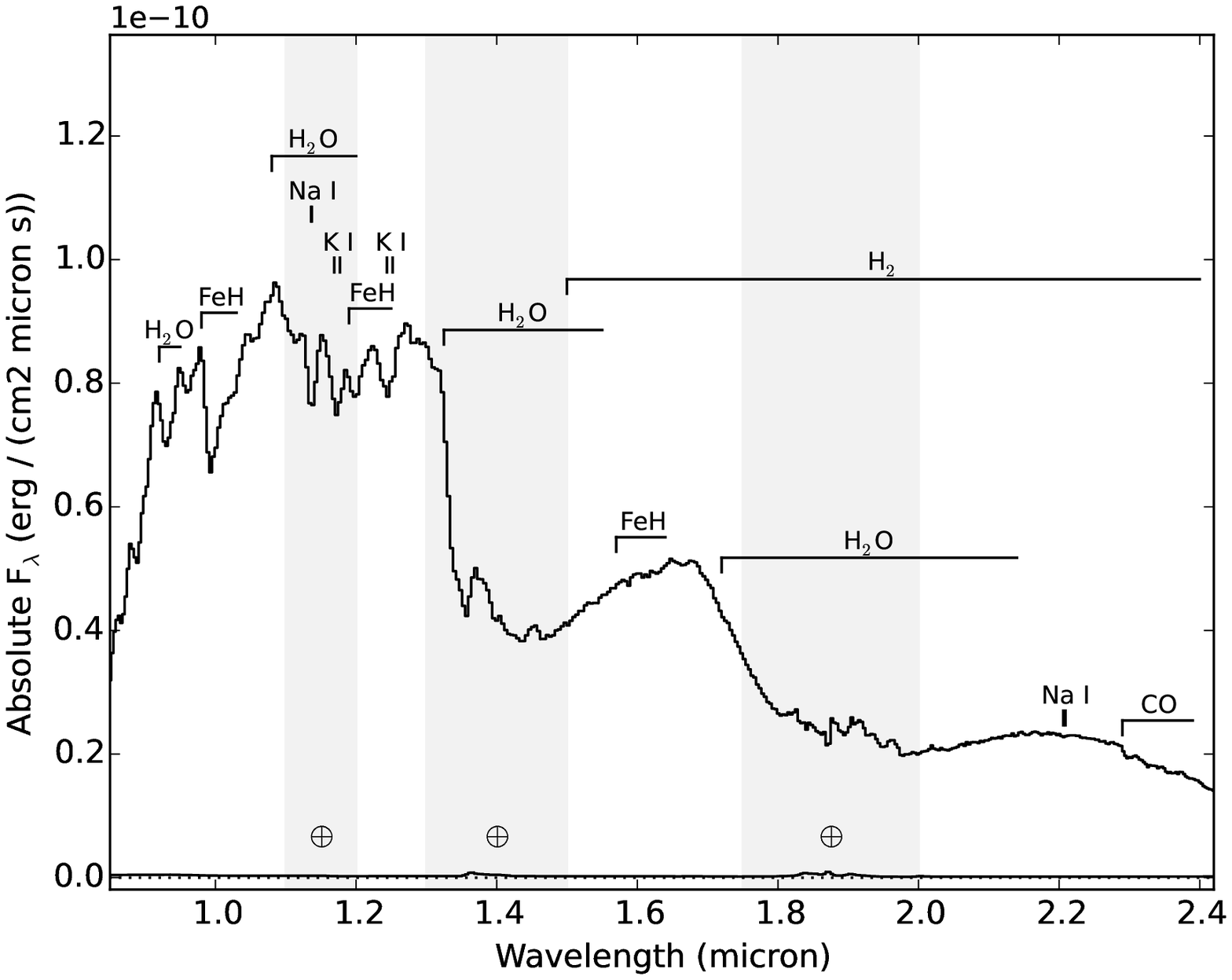}
\caption{ Low resolution SpeX spectra of GJ~660.1A (top) and GJ~660.1B (bottom) with uncertainties. 
The spectra are normalized to their absolute magnitudes.  Spectral features arising from  H$_2$O (0.92~$\micron$, 1.325~$\micron$, 1.72$\micron$), CO (2.3~$\micron$), FeH (1.0~$\micron$, 1.2~$\micron$)  and H$_2$ molecules and various metal lines are labeled. Grey bands mark regions of strong telluric absorption.}
\label{fig:spectra}
\end{figure}

\begin{figure*}
\epsscale{1.0}
\plotone{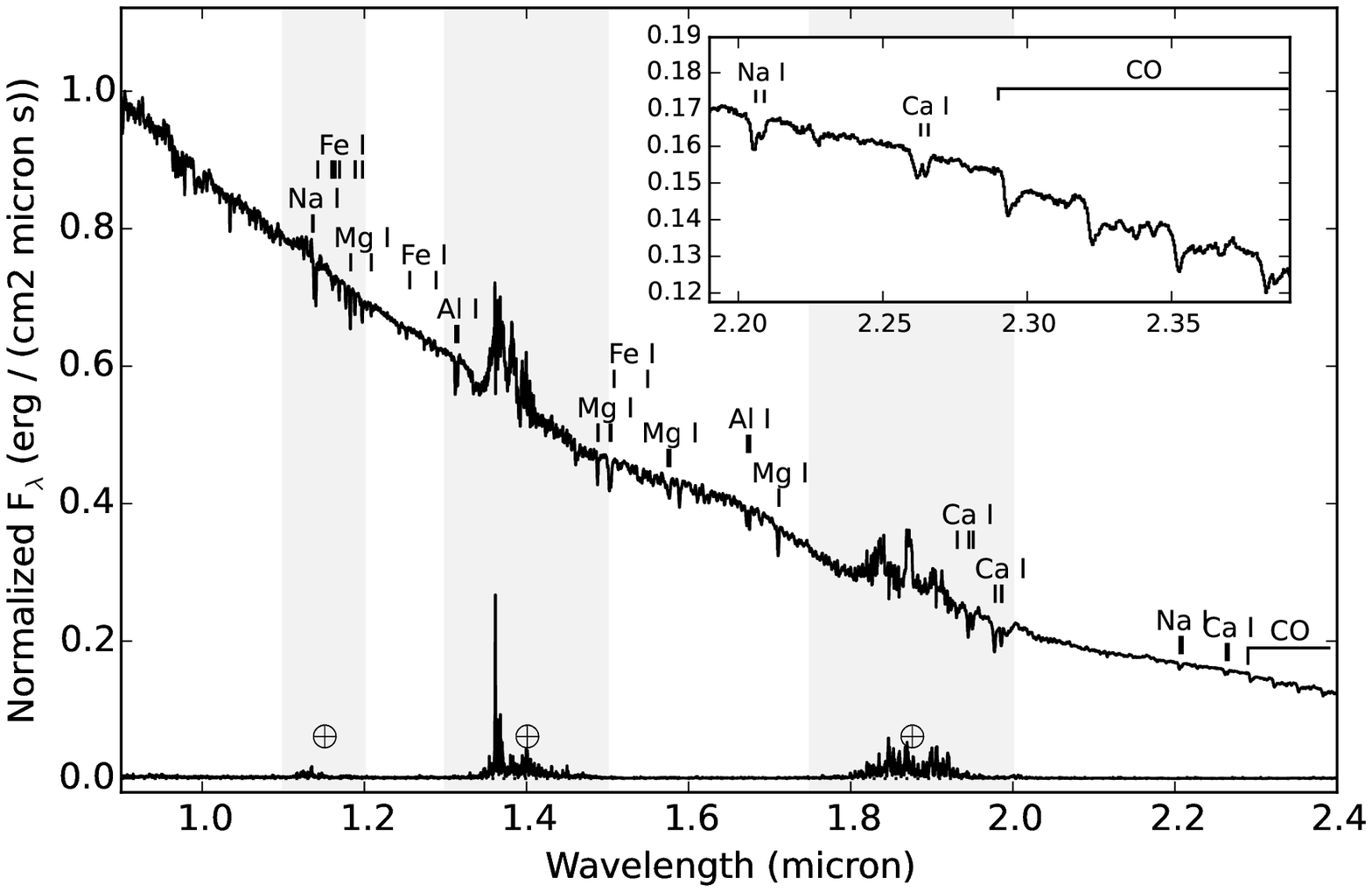}
\caption{ Moderate resolution SpeX SXD spectrum of GJ~660.1A, with uncertainty plotted. 
Spectral features from 
Na I (1.14~$\micron$, 2.21~$\micron$), 
Fe I (1.14--1.20~$\micron$, 1.51~$\micron$, 1.55~$\micron$),
Mg I (1.19~$\micron$, 1.50~$\micron$, 1.58~$\micron$, 1.71~$\micron$), 
Al I (1.31~$\micron$, 1.68~$\micron$) and 
Ca I (1.94~$\micron$, 1.98~$\micron$, 2.26~$\micron$) lines, and CO molecular absorption from 2.28--2.40~$\micron$, are labeled. The inset plot shows a close-up of the 2.0--2.4~$\micron$ region to show the Na I, Ca I and CO features. Grey bands mark regions of strong telluric absorption.}
\label{fig:sxd}
\end{figure*}

The reduced prism spectra of  GJ~660.1A and B  are displayed in Figure~\ref{fig:spectra}, while the reduced SXD spectrum of GJ~660.1A is shown in Figure~\ref{fig:sxd}.  The spectra of the primary show features consistent with its M1 optical classification, including H$_2$O (0.92--0.95~$\micron$, 1.08--1.20~$\micron$, 1.325--1.450~$\micron$ and 1.72--2.14~$\micron$) and CO (2.3~$\micron$) absorption, and numerous atomic metal lines in the moderate-resolution data \citep{2001ApJ...548..908L,2009ApJS..185..289R}. We analyze this spectrum in further detail in Section~\ref{sec:metal-poor}.  The secondary spectrum exhibits features consistent with late M/early L dwarfs, including absorption from  H$_{2}$O, CO and FeH molecules (1.0~$\micron$ and 1.2~$\micron$), and several unresolved alkali lines including K I (1.25~$\micron$) and Na I (1.51~$\micron$; \citealt{2001AJ....121.1710R,2001ApJ...548..908L}). This confirms the photometric classification of \citet{2011ApJ...743..109S}. There are a few clear peculiarities in the spectrum of this source, including its triangle-shaped H-band peak and shallow CO absorption, that are discussed in detail below.  

\section{Spectral Analysis}

\subsection{Classification of GJ~660.1B }

To refine the classification of GJ~660.1B, we first considered the spectral indices and index-SpT relations from \cite{2007ApJ...657..511A}. All measured indices had uncertainties estimated from spectral noise through Monte Carlo techniques, and these were propagated into the spectral type determinations along with each relation's intrinsic scatter. These measurements (Table~\ref{tab:indices}) resulted in a classification of M9.5$\pm$0.3, driven primarily by the strong H$_2$O absorption present in the spectrum of this source.

\begin{deluxetable}{lcc}
\tabletypesize{\small}
\tablecaption{Index Measurements for GJ~660.1B\label{tab:indices}}
\tablewidth{0pc}
\tablehead {
\colhead{Index} & \colhead{Value} & \colhead{Spectral Type} \\
}
\startdata
H2O & 1.119$\pm$0.007 & M8.9$\pm$0.4 \\
H2O-1 & 0.604$\pm$0.003 & L2.8$\pm$1.1 \\
H2O-2 & 0.886$\pm$0.009 & M9.3$\pm$0.6 \\
H2OD & 0.987$\pm$0.011 & [L0.3$\pm$0.8]\tablenotemark{a} \\
\hline
{Mean} & & M9.5$\pm$0.3 \\
\enddata
\tablecomments{Indices defined in \citet{2004ApJ...610.1045S}; \citet{2003ApJ...596..561M} and \citet{2007ApJ...657..511A}.}
\tablenotetext{a}{Outside defined range.}
\end{deluxetable}

\begin{figure}
\epsscale{0.75}
\plotone{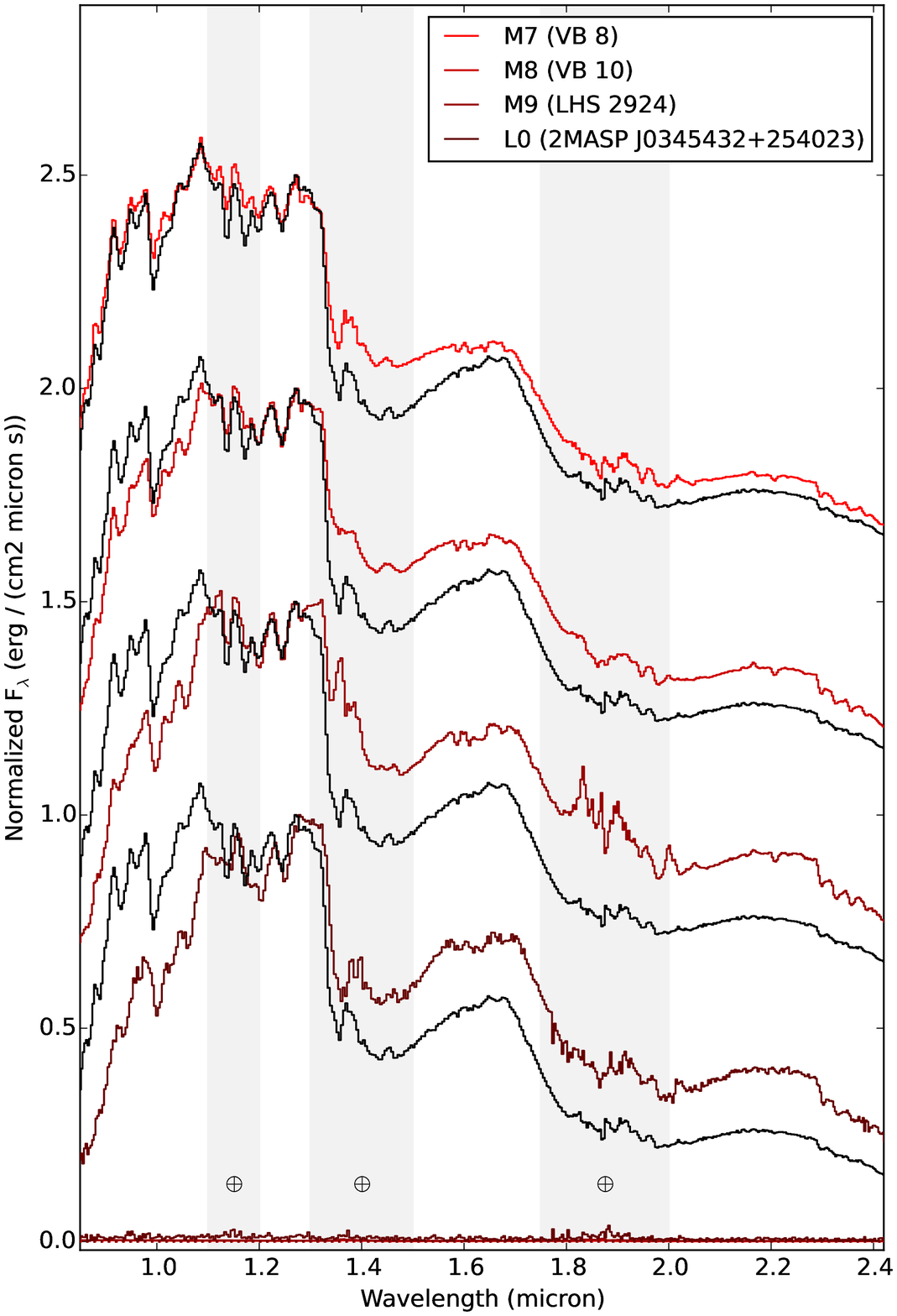}
\caption{Spectrum of GJ~660.1B (black) compared to M7, M8, M9 and L0 spectral standards defined by \citet[descending]{2010ApJS..190..100K}. Standard spectral data are from \citet{2006AJ....131.1007B} and \citet{2014ApJ...794..143B}. All spectra are normalized at their peak flux and offset in steps of 0.5. The M7 standard is the best-fit in the 0.9--1.4~$\micron$ range, but poorly matches the full SED of GJ~660.1B. Grey bands mark regions of strong telluric absorption.}
\label{fig:classification}
\end{figure}

We then compared the spectrum of GJ~660.1B directly to the M and L dwarf spectral standards defined in \citet{2010ApJS..190..100K} using a chi-square statistic:
\begin{equation}
\chi ^{2}(k)=\sum_{\lambda}\ \frac{(x_{\lambda}-\alpha m_{\lambda}(k))^2}{{\sigma}_{\lambda}^{2}}      
\label{eqn:chi2}
\end{equation}
where $x_{\lambda}$ is the measured spectral flux density of GJ~660.1B, $\sigma_{\lambda}$ is the uncertainty of this spectrum, $m_{\lambda}(k)$ is the spectrum of standard $k$, and $\alpha$ is the optimal scaling factor computed as 
\begin{equation}
\alpha =\frac{\sum_{\lambda} x_{\lambda}m_{\lambda}(k)/\sigma _{\lambda}^2}{\sum_{\lambda} m_{\lambda}^2(k)/\sigma _{\lambda}^2}  
\label{eqn:alpha}
\end{equation}
(see \citealt{2008ApJ...678.1372C}).  Following the procedure described in  \citet{2010ApJS..190..100K}, all sums were performed over the wavelength bins spanning 0.9--1.4~$\micron$. The spectral standard with the lowest $\chi^{2}$ was found to be the M7 dwarf VB~8 \citep{1961AJ.....66..528V}, but the agreement is poor beyond 1.35~$\micron$ (Figure~\ref{fig:classification}). The M8, M9 and L0 spectral standards are also poor matches to the spectrum of GJ~660.1B across the {0.8--2.4~$\micron$} near-infrared band.

\begin{figure}
\epsscale{0.75}
\plotone{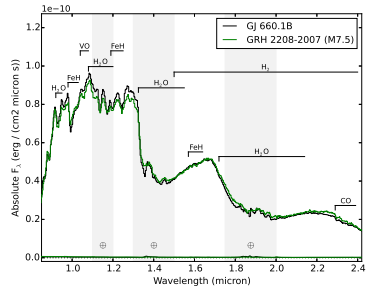}
\caption{Spectrum of GJ~660.1B (black) compared to its best-match, optically-classified spectrum in the SPL, the M7.5 GRH~2208-2007 (green; data from \citealt{2014ApJ...794..143B}).  The spectrum of GJ~660.1B is scaled to absolute fluxes while the comparison spectra are scaled to its optimal scaling factor (Eqn~\ref{eqn:alpha}).
Key absorption features are labeled; grey bands mark regions of strong telluric absorption.}
\label{fig:comp2208}
\end{figure}

Finally, we compared the spectrum of GJ~660.1B to 911 optically-classified M5--L5 spectra in the SpeX Prism Library (SPL; \citealt{2014ASInC..11....7B}) using the same fit statistic as above. Comparisons were made over the wavelength ranges 0.7--1.35~$\micron$, 1.42--1.8$\micron$, and 1.92--2.45~$\micron$ to avoid regions of strong telluric absorption. The best-fit template using the same $\chi^2$ statistic above is the high proper-motion ($\vec{\mu}$ = [$-$124$\pm$50, $-$821$\pm$8]~mas~yr$^{-1}$) M7.5 brown dwarf GRH~2208-2007 \citep[Figure~\ref{fig:comp2208}]{1998A&A...338.1066T,2003AJ....126.2421C,2005A&A...435..363D} which fits significantly better than VB~8, although GJ~660.1B has a somewhat bluer overal spectral energy distribution (SED).

In summary, we found significant discrepancies in the classifcation of GJ~660.1B between three methods, with results ranging from M7 (standards) to M9.5 (indices). More importantly, the spectrum deviates significantly from the spectral standards in this type range.  A robust classification of this source requires more detailed examination of these peculiarities.

\subsection{ Peculiarities in the Spectrum of GJ~660.1B }

\subsubsection{ Hypothesis 1: A Young Brown Dwarf }

The spectrum of GJ~660.1B exhibits an $H$-band (1.7~$\micron$) peak that resembles those of many low surface gravity brown dwarfs in young clusters and associations
(e.g., \citealt{2001MNRAS.326..695L,2006ApJ...639.1120K,2007ApJ...657..511A,2013ApJ...772...79A}). This spectral region in M and L dwarfs is shaped by H$_2$O and FeH molecular bands and H$_2$ collision-induced absorption (CIA; \citealt{1969ApJ...156..989L,1997A&A...324..185B,2008ApJ...674..451B,2013ApJ...772...79A}), the last of which is a pressure-sensitive opacity source. Its triangle-shaped $H$-band peak suggests that GJ~660.1B may have a low surface gravity.  To quantify this hypothesis, we used the index-based gravity classification scheme of \citet{2013ApJ...772...79A}, which examines gravity-sensitive features from VO, FeH, alkali lines and continuum shape. For a classification of M9.5, the resulting gravity scores (Table~\ref{tab:gravity}) yield a classification of INT-G (intermediate low surface gravity), which is consistent with objects of ages $\approx$100--300 Myr \citep{2013ApJ...772...79A}.  However, if we assume a spectral type of M7, the scores instead yield a FLD-G (field dwarf) gravity classification.

\begin{deluxetable}{lccc}
\tabletypesize{\small}
\tablecaption{Gravity Score Measurements for GJ~660.1B\label{tab:gravity}}
\tablewidth{0pc}
\tablehead {
\colhead{Index} & \colhead{Value} & \colhead{Gravity} & \colhead{Gravity} \\
 & & \colhead{Score (M9.5)} &  \colhead{Score (M7)} \\
}
\startdata
VO-z & 1.004$\pm$0.004 &  -- & -- \\
FeH-z & 1.189$\pm$0.006 & 0 & 0 \\
H-cont & 0.961$\pm$0.006 & 1 & 0 \\
KI-J & 1.084$\pm$0.003 & 1 & 0 \\
\hline
{Mean} & & INT-G & FLD-G \\
\enddata
\tablecomments{Indices defined in \citet{2013ApJ...772...79A}.}
\end{deluxetable}

\begin{figure}
\epsscale{0.75}
\plotone{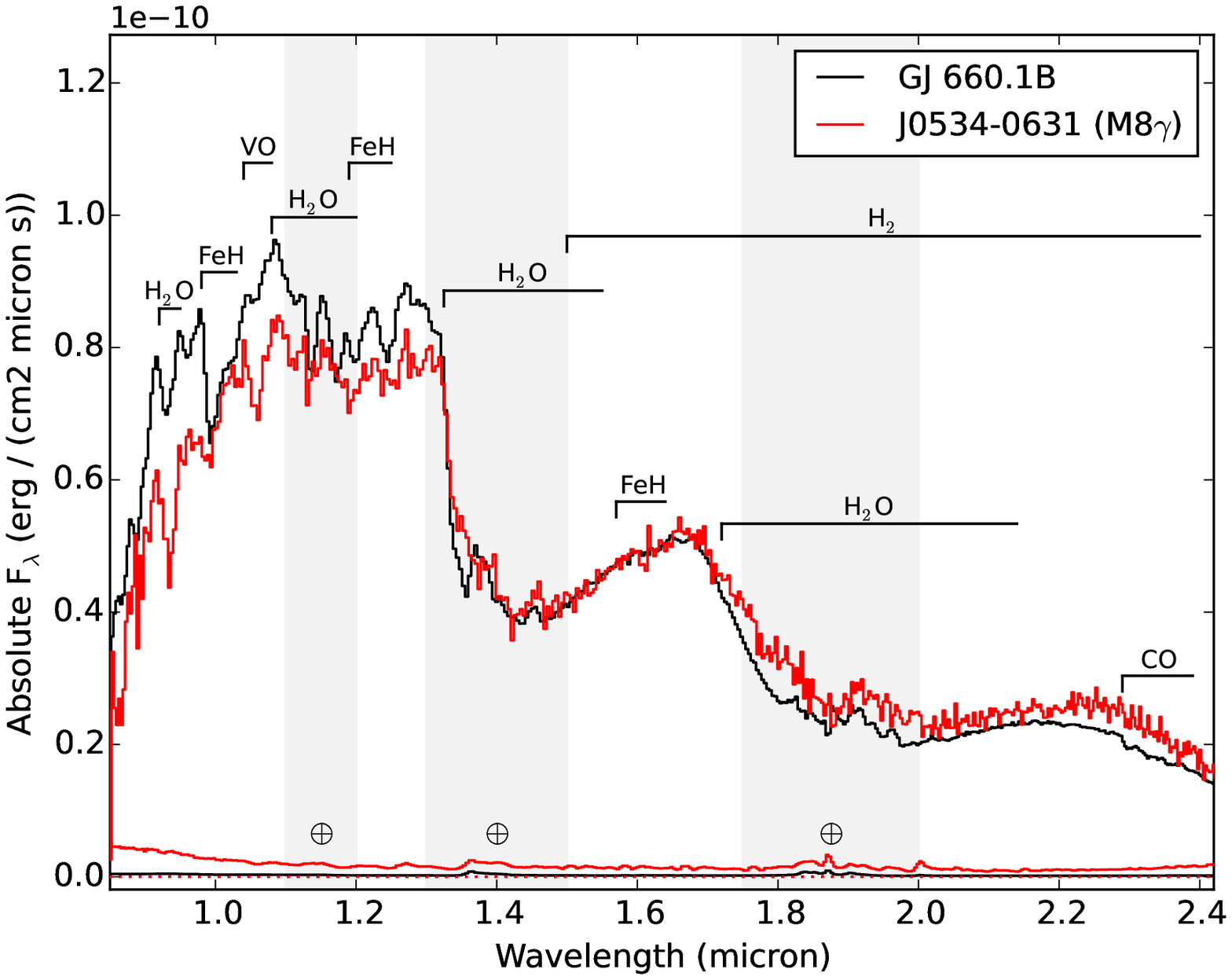}
\plotone{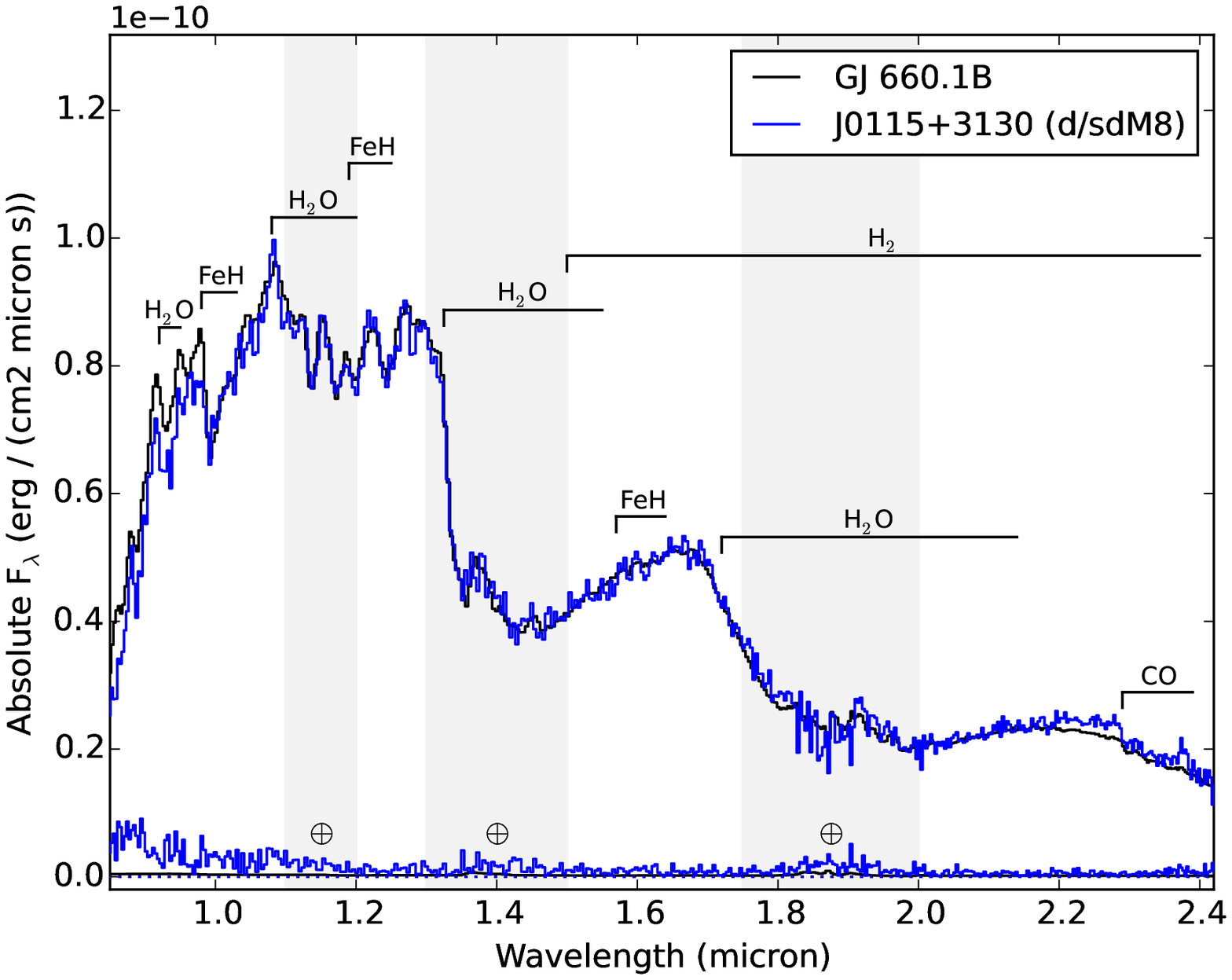}
\caption{Same as Figure~\ref{fig:comp2208}, comparing the spectrum of GJ~660.1B (black) to its best-match young brown dwarf, the M8$\gamma$ 2MASS J05341592$-$0631401 (top in red; data from \citealt{2010ApJS..190..100K}) and its best-match metal-poor dwarf, the d/sdM8 2MASS J01151621+3130061 (bottom in blue; data from \citealt{2004AJ....127.2856B}).}
\label{fig:comparison}
\end{figure}

To assess {these conflicting results}, we compared the spectrum of GJ~660.1B directly to other low- and intermediate-gravity spectra in the SPL. The best-matching case is the M8$\gamma$\footnote{The $\gamma$ suffix is assigned for ages $\approx$10--30~Myr \citep{2009AJ....137.3345C,2010ApJS..190..100K}, although 2MASS J05341592$-$0631401 has not yet been assigned to a particular association or moving group (J.\ Gagne, 2015, private communication).}/M8 VLG 2MASS J05341592$-$0631401 (\citealt{2010ApJS..190..100K,2013ApJ...772...79A}; Figure~\ref{fig:comparison}). While the $H$-band peaks are similar between these sources, there is a clear mismatch in the 0.9--1.4~$\micron$ region. While 2MASS J05341592$-$0631401 shows weak FeH absorption at 1.0~$\micron$ and strong VO at 1.05~$\micron$, the spectrum of GJ~660.1B has strong FeH and no discernable VO band.  In addition, the overall SED of GJ~660.1B is significantly bluer ($J-K_s$ = 0.82$\pm$0.06) than comparably classified dwarfs 
($\langle{J-K_s}\rangle$ = 1.08$\pm$0.19 for M7 dwarfs, $\langle{J-K_s}\rangle$ = 1.20$\pm$0.22 for M9 dwarfs; \citealt{2009AJ....137....1F}), whereas young brown dwarfs tend to be 0.1--0.3 magnitudes redder \citep{2012ApJ...752...56F}. We therefore conclude that GJ~660.1B is not young, and that its peculiarities arise from another source.

\subsubsection{ Hypothesis 2: A Metal-Poor Dwarf \label{sec:metal-poor} }

The best-matching SPL spectra to GJ~660.1B have common physical properties. Two particularly good matches are 2MASS~J01151621+3130061, a moderately metal-poor d/sdM8 dwarf \citep[Figure~\ref{fig:comparison}]{2009ApJ...706.1114B}; and HD~114762B, a d/sdM9$\pm$1 which is a companion to a metal-poor star ([Fe/H] = $-$0.7; \citealt{1989Natur.339...38L,1995PASP..107...22H,2009ApJ...706.1114B}). The d/sd ``mild subdwarf'' classifications of these sources reflect spectral peculiarities arising from their slightly sub-solar metallicities \citep{2007ApJ...657..494B}.  These spectra show the same triangular-shaped $H$-band peaks, strong FeH and alkali line absorption, and blue SEDs as GJ~660.1B.

To validate this hypothesis, we used the moderate-resolution near-infrared SpeX spectrum of GJ~660.1A (Figure~\ref{fig:sxd}) to re-assess its metallicity. As described in \cite{2010ApJ...720L.113R,2012ApJ...748...93R}, metal lines in the near-infrared can be used to determine the metallicities of M-type dwarfs based on empirical calibrations from
M dwarf companions to FGK stars. {We used the metallicity calibration relations defined by
 \citet{2012ApJ...748...93R}, \citet{2012ApJ...747L..38T}, \citet{2013AJ....145...52M}, and \citet{2014AJ....147...20N}, which are based on the equivalent widths of the 1.516~$\micron$ K~I, 2.205~$\micron$ Na~I, and 2.263~$\micron$ Ca~I metal lines and the H$_2$O spectral indices defined in \citet{2010ApJ...722..971C} and \citet{2012ApJ...748...93R}.  The resulting estimates of [Fe/H] and [M/H] are listed in Table~\ref{tab:metallicity}.  All of the estimates are in agreement except those from \citet{2012ApJ...747L..38T}, which exhibit a large difference between $H$- and $K$-band calibrations.  Rejecting these, we determine mean values of [Fe/H] = $-$0.63$\pm$0.06 and [M/H] = $-$0.47$\pm$0.07 for GJ~660.1A. The former is lower than but statistically consistent with [Fe/H] = $-$0.88$\pm$0.13 as reported by 
\citet{2014MNRAS.443.2561G}, based on the calibration of optical metal lines given by \citet{2013AJ....145...52M}.
We adopt the near-infrared metallicity determination for GJ~660.1A the remainder of this study.}

\begin{deluxetable*}{lcc}
\tabletypesize{\small}
\tablecaption{Metallicity Determinations for GJ~660.1A\label{tab:metallicity}}
\tablewidth{0pc}
\tablehead {
\colhead{Method} & \colhead{[Fe/H]} & \colhead{[M/H]} \\
}
\startdata
\citet{2012ApJ...748...93R} $K$-band & $-$0.57$\pm$0.14 & $-$0.40$\pm$0.10 \\
\citet{2012ApJ...747L..38T} $H$-band & $-$0.18$\pm$0.12\tablenotemark{a} & \nodata \\
\citet{2012ApJ...747L..38T} $K$-band & $-$1.08$\pm$0.12\tablenotemark{a} & \nodata \\
\citet{2013AJ....145...52M} $J$-band & $-$0.60$\pm$0.07 & $-$0.32$\pm$0.08 \\
\citet{2013AJ....145...52M} $H$-band & $-$0.62$\pm$0.08 & $-$0.52$\pm$0.06 \\
\citet{2013AJ....145...52M} $K$-band & $-$0.64$\pm$0.06 & $-$0.50$\pm$0.05 \\
\citet{2014AJ....147...20N} $K$-band & $-$0.80$\pm$0.13 & \nodata \\
\hline
{Weighted Mean} & $-$0.63$\pm$0.06 & $-$0.47$\pm$0.07 \\
Optical \citep{2014MNRAS.443.2561G} & $-$0.88$\pm$0.13 & \nodata \\
\enddata
\tablenotetext{a}{Not included in uncertainty-weighted mean.}
\end{deluxetable*}

\begin{figure}
\epsscale{0.75}
\plotone{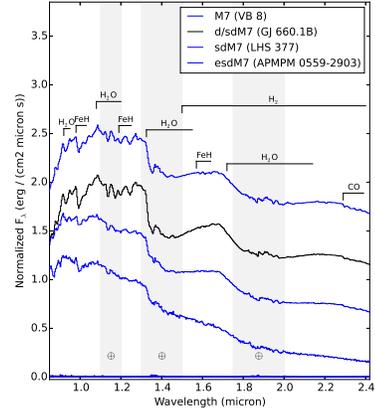}
\caption{\bf Near-infrared spectrum of GJ~660.1B (black line) compared to (blue lines)
the M7 VB~8 (data from \citealt{2014ApJ...794..143B}),
the sdM7 LHS~377 (data from \citealt{2004ApJ...614L..73B}),
and the esdM7 APMPM~0559-2903 (data from \citealt{2006ApJ...645.1485B}).
All spectra are normalized to their 1.25--1.30~$\micron$ peaks and offset by constats.
Grey bands mark regions of strong telluric absorption; key molecular features are labeled.}
\label{fig:sdcomp}
\end{figure}

{The proximity and co-movement of the two components of GJ~660.1AB strongly indicate coevality, and} the subsolar metallicity {of the primary} supports our second hypothesis that the spectral peculiarities seen in GJ~660.1B {likely} arise from low metallicity effects. 
{These peculiarities are not as extreme as those seen in M subdwarfs and extreme M subdwarfs, 
as illustrated in Figure~\ref{fig:sdcomp}, which shows that the near-infrared spectral morphology of GJ~660.1B is intermediate between those of M7 and sdM7 dwarfs.}
Following \citet{2009ApJ...706.1114B} and \citet{2010ApJS..190..100K}, we assign a classification of d/sdM7 for this source based on its similarity to the M7 standard over 0.9--1.4~$\micron$ and {mild} low metallicity features across the full SED
(see also \citealt{2008AJ....136..840J}). 

\subsubsection{Hypothesis 3: An Unresolved Binary \label{sec:binary} }

\begin{figure}
\epsscale{0.75}
\plotone{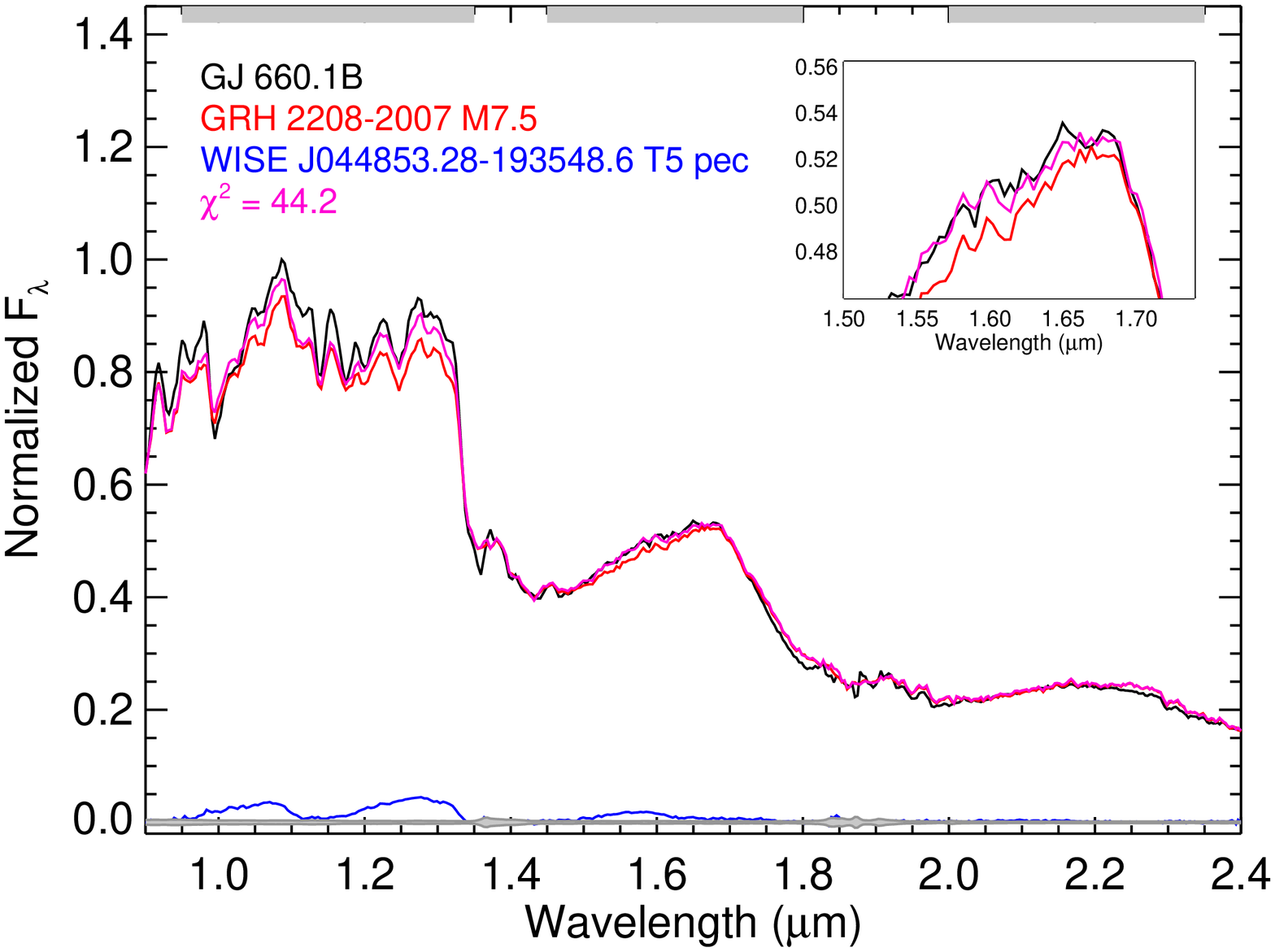}
\caption{Comparison of the low resolution SpeX spectrum of GJ~660.1B (black) to its best-fit SPL binary template (purple) constructed from the spectra of GRH~2208-2007 (red) and the T5 pec dwarf WISE~J044853.28$-$193548.6 (blue; \citealt{2011ApJS..197...19K}).  The uncertainty spectrum is shown at bottom in grey, while comparison regions (1.00--1.35~$\micron$, 1.45--1.80~$\micron$ and 2.00--2.35~$\micron$) are indicated as shaded regions at top. The inset box shows a close-up of the fit in the 1.50--1.75~$\micron$ region. }
\label{fig:binary}
\end{figure}

{We also considered a third hypothesis, that the unusual spectral features of GJ~660.1B arise from the spectral contamination of an unresolved binary (or in this case, tertiary) companion. This scenario has used to explain peculiar spectral features in several late M dwarfs which are suspected or have been confirmed to host T dwarf companions (e.g., \citealt{2008ApJ...681..579B,2012ApJ...757..110B,2015AJ....149..104B,2014ApJ...794..143B}).  To examine this hypothesis, we followed the methods described in \citet{2010ApJ...710.1142B} and \citet{2014ApJ...794..143B}, comparing the spectrum of GJ~660.1B to 487 M6--L2 dwarfs not previously classified as metal-poor, young or binary in the SPL. We then constructed 279,237 synthetic binary template spectra from these primaries and 574 L0--T6 dwarfs (assuming the secondary spectral type is the same or later than the primary type), using the \citet{2012ApJS..201...19D} $M_J$ versus spectral type relation to scale the individual component spectra. Figure~\ref{fig:binary} displays the best-fit binary templates resulting from this analysis, which is a combination of GRH~2208-2007 and the  T5 pec dwarf WISE~J044853.28$-$193548.6 \citep{2011ApJS..197...19K}. While the binary template is a better fit ($\chi^2$ = 44.2 versus 75.2), it fails to surpass the 99\% significance threshold in an F-test comparison, and fails to reproduce structure in the 1.0--1.2~$\micron$ and 2.2--2.3~$\micron$ range. We therefore conclude that unresolved multiplicity is not the most likely source for the observed spectral peculiarities of GJ~660.1B.}

\subsection{  Model Fitting }

\subsubsection{ Methodology }

To better quantify the physical parameters of GJ~660.1B, we compared its spectrum  to two sets of atmosphere models characterized by {\teff}, {\logg} and [M/H]: the BT-Settl models (\citealt{2011ASPC..448...91A}) and the Drift models \citep{2008ApJ...675L.105H,2009A&A...506.1367W,2011A&A...529A..44W}. For the former, we sampled models spanning effective temperatures {\teff} = 1400--2900~K, {\logg} (in cm~s$^{-2}$) = 3.5--5.5, and [M/H] = $-$4.0 to +0.5; for the latter, we sampled {\teff} = 1700--3000~K, {\logg} = 5.0--5.5, and [M/H] = $-$3.0 to 0.0. 
In both cases, the models were originally calculated every 100 K in {\teff}, 0.5 dex in {\logg} and 0.5 dex in [M/H], and at much higher resolution than the SpeX data. We smoothed the models to {\ldl} = 120 using a Hann filter and interpolated onto a common wavelength scale. As the models are calculated in units of surface flux densities, we calibrated the observed spectrum of GJ~660.1B to absolute flux densities using its absolute 2MASS $J$-band magnitude ($M_J$ = 11.54$\pm$0.16; \citealt{2011ApJ...743..109S}). The optimal scaling factor (Eqn~\ref{eqn:alpha}) is then $\alpha$ = $F_{\rm spectrum}$/$F_{\rm model}$ = (R/10~pc)$^{2}$, where R is the radius of GJ~660.1B. 
These fits therefore provide an estimate of the source size \citep{2009ApJ...706.1114B}.

We deployed a Markov Chain Monte Carlo  (MCMC) code with a Metropolis-Hastings algorithm \citep{1953JChPh..21.1087M,HASTINGS01041970} to identify the best-fit model parameters and uncertainties for both sets of atmosphere models. We used initial estimates of {\teff} = 2650 K based on its d/sdM7 spectral type and the {\teff}/spectral type relations of \citet{2009ApJ...702..154S}; \citet{2013AJ....146..161M} and \citet{2015arXiv150801767F}. We also assumed {\logg} = 5.0, which is typical for a field late M dwarf; and [M/H] = 0.0. Two 10$^5$-step chains were run in which our three model parameters were alternately updated using step sizes drawn from normal distributions of width $\Delta${\teff} = 50~K, $\Delta${\logg} = 0.25, and  $\Delta$[M/H] = 0.25. Models at intermediate parameter values were linearly interpolated in logrithmic flux and compared to the observed spectrum using the same $\chi^2$ statistic as Eqn~\ref{eqn:chi2}. 
The criterion to adopt a successive parameter $\theta_i \rightarrow \theta_{i+1}$ was $U(0,1) \leq e^{-0.5(\chi^2_{(i+1)}-\chi^2_{(i)})}$, where $U(0,1)$ is a random number drawn from a uniform distribution between 0 and 1.  We eliminated the first 10\% of each chain before evaluation of the model parameters. 

\subsubsection{ Results }

\begin{figure}
\epsscale{1.0}
\plotone{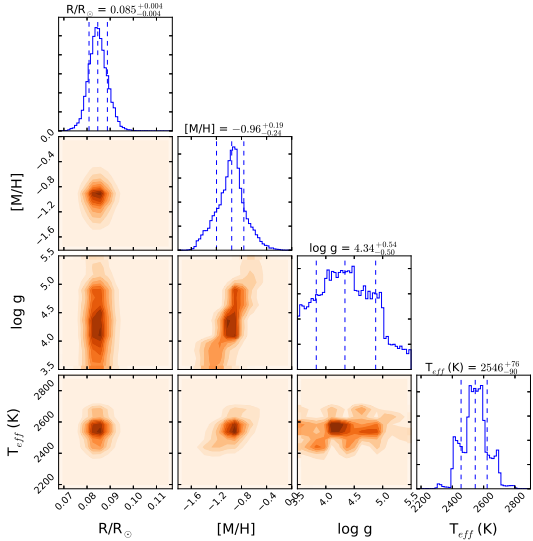}
\caption{{\teff}, {\logg}, [M/H] and radius distributions based on MCMC model fitting of the SpeX spectrum of GJ~660.1B using BT-Settl models. Color contours display the density of model parameters sampled in our MCMC chain for the corresponding parameter pairs. These show a strong positive correlation between {\logg} and [M/H].  One-dimensional histograms are marginalized over all other parameters, with median and 16\% and 84\% quantiles labeled and listed. Note that structure in the {\teff} distribution arises from our model interpolation scheme. This plot was generated using code by \citet{dan_foreman_mackey_2014_11020}.}
\label{fig:parameters_btsettled}
\end{figure}

\begin{figure}
\epsscale{1.0}
\plotone{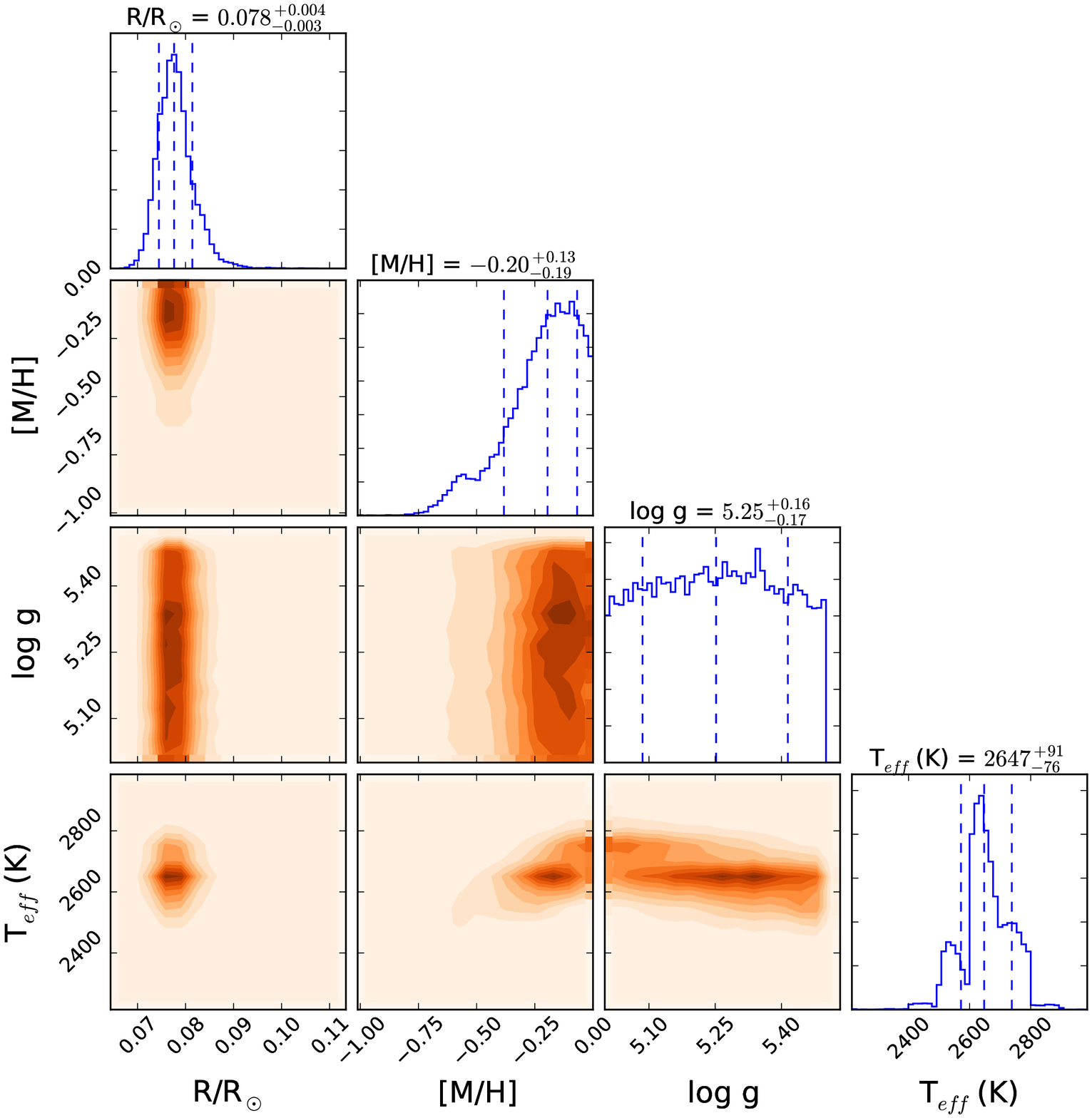}
\caption{Same as Figure~\ref{fig:parameters_btsettled} for Drift models.}
\label{fig:parameters_drift}
\end{figure}

\begin{figure}
\epsscale{0.75}
\plotone{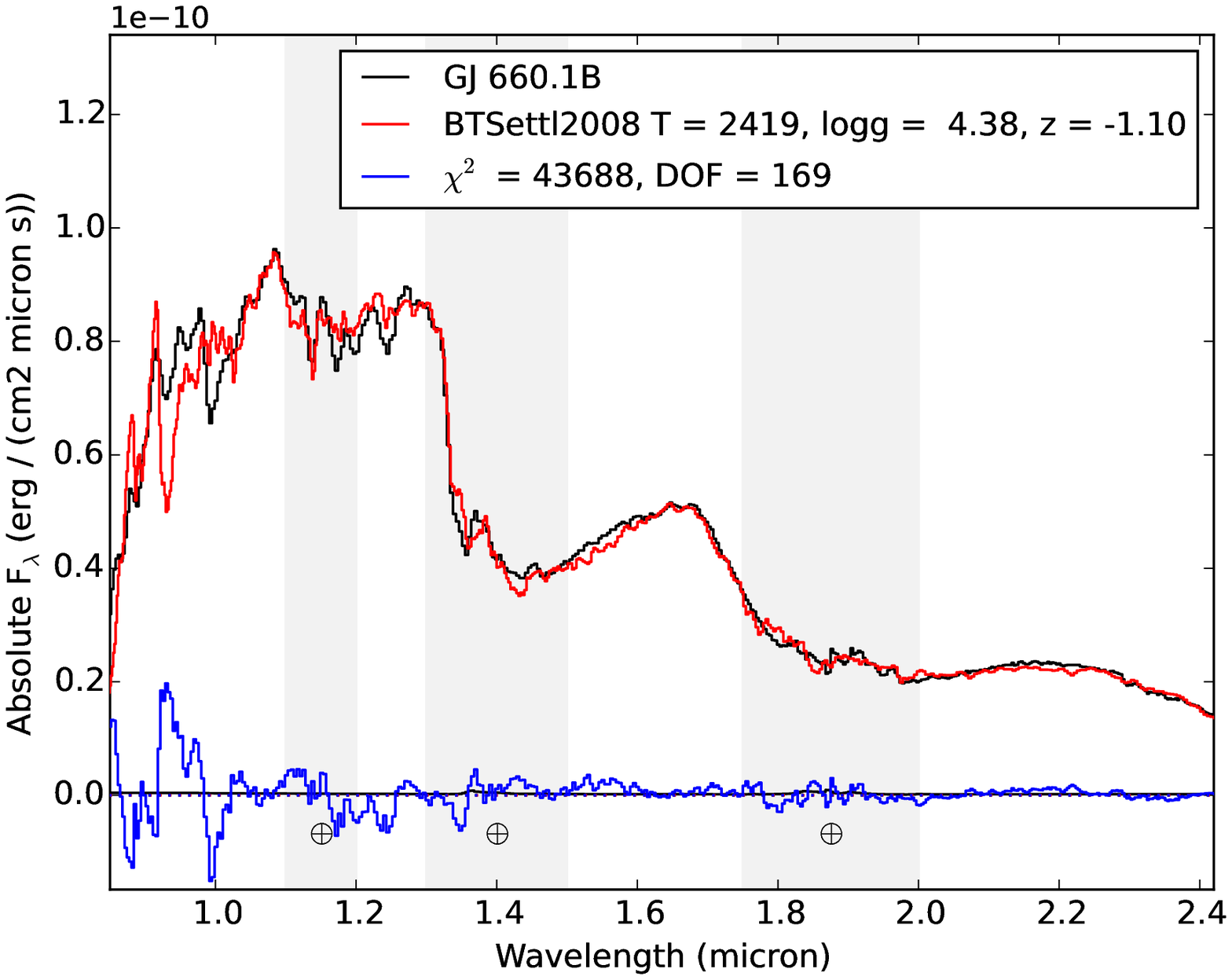}
\plotone{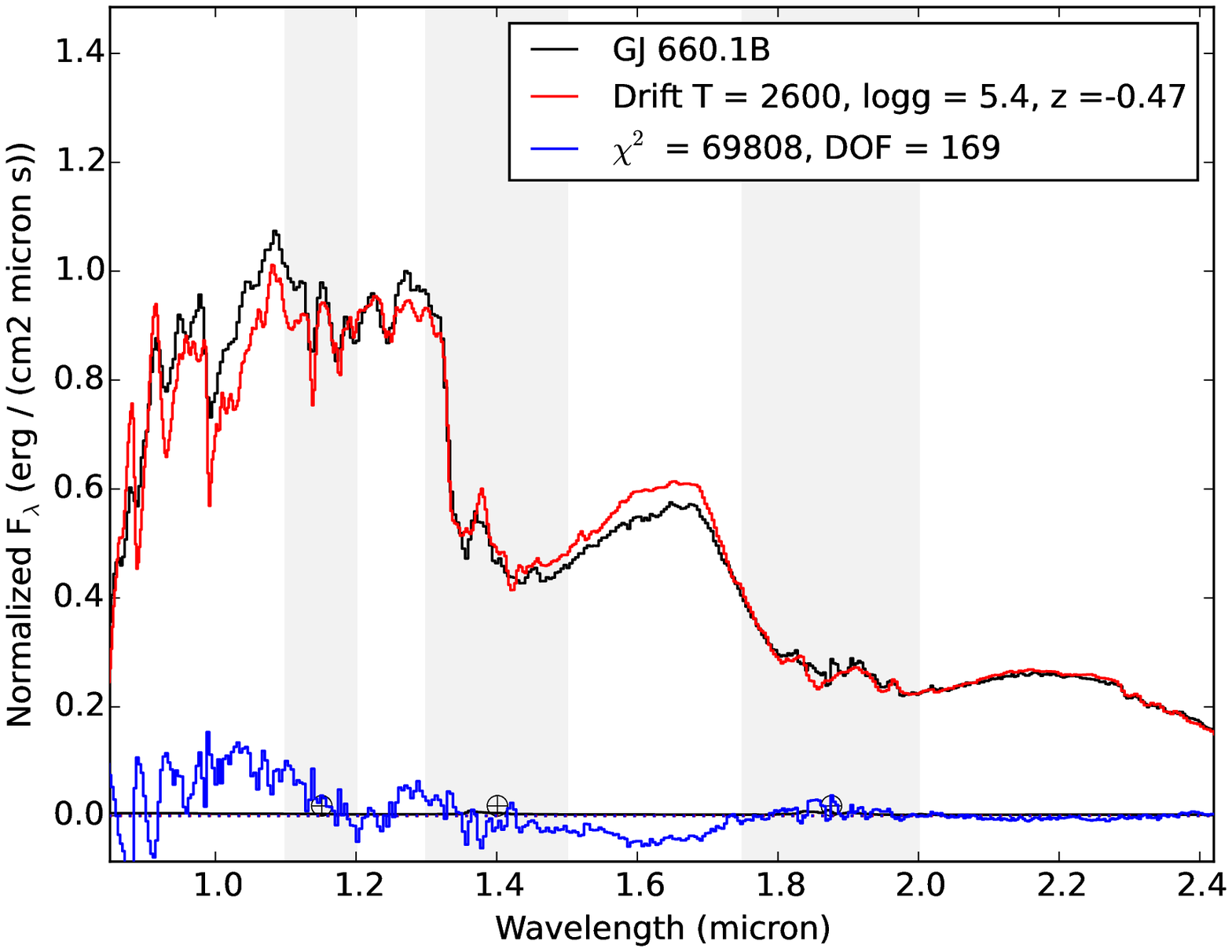}
\caption{Comparison of the observed spectrum of GJ~660.1B (black lines) to BT-Settl models (red lines). The top panel shows the best-fit model parameters, the bottom panel shows an ``optimal'' parameter set of {\teff} = 2600~K, {\logg} = 5.4 and [M/H] = {$-$0.47}.  The spectrum of GJ~660.1B is scaled to its absolute 2MASS $J$ magnitude, while the model spectra are scaled to their optimal scaling factors (Eqn.~\ref{eqn:alpha}).  The difference spectra (GJ~660.1B - Model) are shown as blue lines. The inset box lists the corresponding $\chi^2$ and degrees of freedom (DOF) for the fits. Grey bands delineate regions of strong telluric absorption not included in the fits.}
\label{fig:bestfit_btsettled}
\end{figure}

The distributions\footnote{Structure in the parameter distributions is primarily an artifact of the model interpolation scheme employed.} of model parameters are shown in  Figures~\ref{fig:parameters_btsettled} and~\ref{fig:parameters_drift}, and summarized in Table~\ref{tab:modelfit}.  For the BT-Settl models, we find median parameters of {\teff} = {\teffresult}~K, {\logg} = {\loggresult}, and [M/H] = {\zresult}, where uncertainties represent the 16\% and 84\% quantiles in the marginalized parameter distributions.  Both {\teff} and [M/H] are well-constrained, the former consistent with the classification of the source. {However, the metallicity is 2.4$\sigma$ lower than the metallicity determined for the GJ~660.1A.} Surface gravity constraints are weaker, and the fits primarily limit {\logg} $\lesssim$ 5.0. As discussed below, this is problematic, as it predicts a very low mass and age for GJ~660.1B, contrary to the comparative analysis above.  There is a strong correlation between [M/H] and {\logg}, with a slope of d[M/H]/d{\logg} $\approx$ 0.35. This likely arises from a trade-off in the strength of H$_2$ CIA, which increases with both increasing surface gravity and decreasing metallicity (e.g., \citealt{2009A&A...506.1367W}).
Solutions with {\logg} $>$ 5 have mean metallicities of $-$0.63$\pm$0.15, which is {more in line with the metallicity of GJ~660.1A}. In contrast, the best-fit model (Figure~\ref{fig:bestfit_btsettled}) has both a low {\logg} = 3.85 and a very low [M/H] = $-$1.20. This model is a reasonable match to the SED of GJ~660.1B beyond 1.4~$\micron$ but poorly reproduces the detailed features in the 0.9--1.4~$\micron$ region, most notably FeH absorption at 1.0~$\micron$ and 1.2~$\micron$. Higher surface gravity and higher metallicity models match the $<$1.4~$\micron$ wavelength end of the spectrum better, but poorly reproduce the full SED.

\begin{figure}
\epsscale{0.75}
\plotone{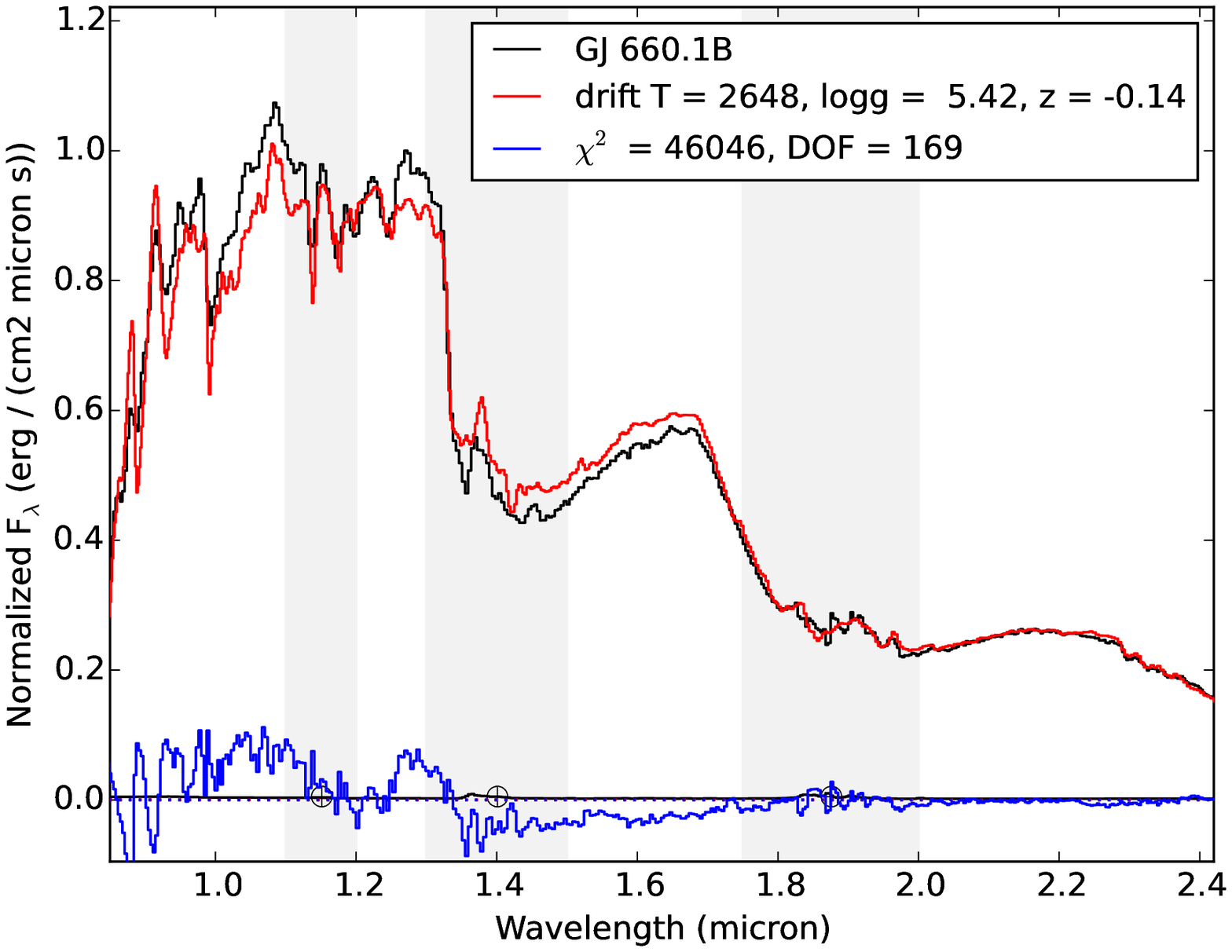}
\plotone{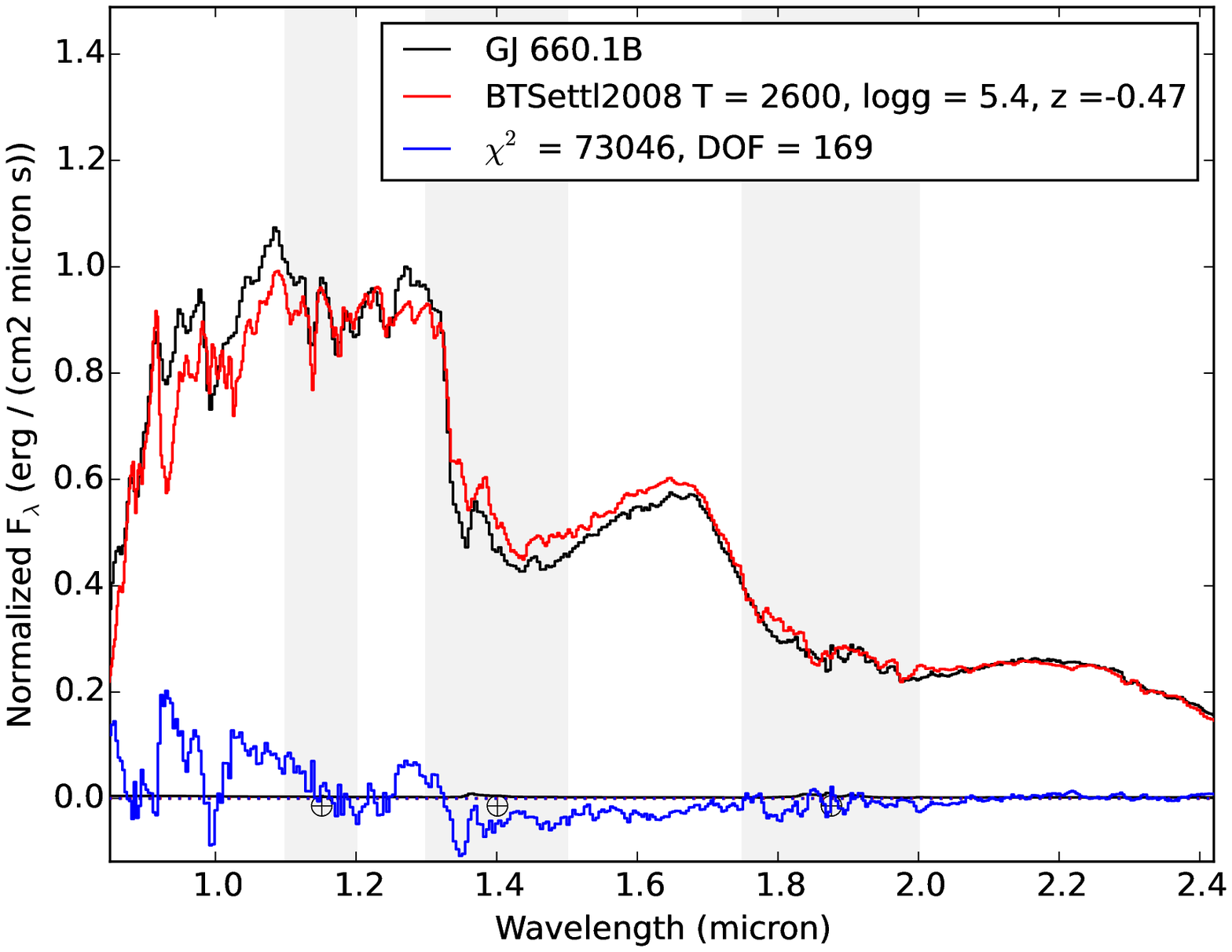}
\caption{Same as Figure~\ref{fig:bestfit_btsettled} for the Drift models.}
\label{fig:bestfit_drift}
\end{figure}

\begin{deluxetable*}{ccccc}
\tabletypesize{\small}
\tablecaption{Model Fitting Results \label{tab:modelfit}}
\tablewidth{0pc}
\tablehead {
\multicolumn{5}{c}{Spectral Models} \\
\cline{1-5}
\colhead{Spectral Model}	& \colhead{$[M/H]$} & \colhead{{\teff} (K)} & \colhead{{\logg} ($cm/s^2$)} & \colhead{Radius ($R_\sun$)} 
}
\startdata
{BT-Settl} &   $-$0.96$^{+0.19} _{-0.24}$ & 2550$^{+80} _{-90}$ & 4.4$^{+0.5} _{-0.5}$ & 0.085$\pm0.014$  \\ 
{Drift} &  $-$0.20$^{+0.13} _{-0.19}$ & 2650$^{+90} _{-80}$ & 5.0--5.5 & 0.078$\pm$0.013 \\
\hline
\\
\multicolumn{5}{c}{Spectral and Evolutionary Models} \\
\hline
\multicolumn{1}{c}{Models} &
\multicolumn{1}{c}{Mass (M$_{\odot}$)} &
\multicolumn{1}{c}{Age (Gyr)} &
\multicolumn{1}{c}{Luminosity (L$_{\odot}$)} &
\multicolumn{1}{c}{Radius (R$_{\odot}$)} \\
\hline
 {BT-Settl \& Baraffe} & 0.038$^{+0.018} _{-0.009}$ & 0.022$^{+0.081} _{-0.011}$& $-$2.8$^{+0.3} _{-0.3}$& 0.23$^{+0.10} _{-0.08}$ \\
 {BT-Settl \& Burrows} & 0.031$^{+0.015} _{-0.006}$ & 0.027$^{+0.065} _{-0.009}$& $-$2.9$^{+0.2} _{-0.3}$& 0.21$^{+0.04} _{-0.07}$ \\
{ Drift \& Baraffe}& 0.088$^{+0.005} _{-0.008}$ & 0.6$^{+0.9} _{-0.4}$ & $-$3.19$^{+0.13} _{-0.10}$ & 0.120$^{+0.017} _{-0.008}$  \\
{  Drift \&  Burrows} & 0.078$^{+0.008} _{-0.009}$& 0.4$^{+0.6} _{-0.2}$ & $-$3.26$^{+0.13} _{-0.10}$ & 0.110$^{+0.015} _{-0.008}$ \\
\enddata
\end{deluxetable*}

The radius inferred from the scale factor is 0.085$\pm$0.014~{\rsun}, where we have incorporated the uncertainty in the distance determination. This is smaller than, but not inconsistent with, radius estimates of 0.10--0.11~{\rsun} based on the evolutionary models \citep{2001RvMP...73..719B,2003A&A...402..701B} for {\teff} = 2550~K and an age $\tau$ = 5~Gyr. However, at this age evolutionary models predict a surface gravity of 5.3--5.4, considerably higher than the surface gravity inferred from the atmosphere model fits.  For the median {\teff} and {\logg} from these fits, the evolutionary models of \citet{2003A&A...402..701B} predict a mass M = 0.038$^{+0.018}_{-0.009}$~{\msun}, $\tau$ = 22$^{+81}_{-11}$~Myr and R$_{evol}$ = 0.23$^{+0.10}_{-0.08}$~{\rsun}, all consistent with a very young brown dwarf, our rejected hypothesis. The \citet{2001RvMP...73..719B} evolutionary models yield similar parameters (Table~\ref{tab:modelfit}).  Note that the evolutionary radius is nearly three times larger (1.8$\sigma$ discrepant) than the scaling factor radius. Bringing these into agreement requires a higher {\logg} in order to lower R$_{evol}$, since the scaling factor is primarily set by the {\teff} of the model (which is uncorrelated with the other parameters) and the observed absolute magnitude of GJ~660.1B. 

For the Drift models, we find an equivalent {\teff} = 2650$^{+90}_{-80}$~K and R = 0.078$\pm$0.013~{\rsun},
but a much higher [M/H] = $-$0.20$^{+0.13}_{-0.19}$, although the metallicity distribution has a long tail toward lower values.
The Drift models were constrained to have 5.0 $\leq$ {\logg} $\leq$ 5.5, and there is no clearly preferred value in that range from the MCMC analysis. 
The Drift metallicity is inconsistent with that derived from the BT-Settl model (2.8$\sigma$ discrepant) and {considerably lower than metallicity of the primary (1.3$\sigma$ discrepant)}, and this may be related to the surface gravity limits on the Drift models. There also appears to be a slight positive correlation between {\teff} and [M/H] and a slight negative correlation between {\teff} and {\logg} in these fits. The best-fit Drift model (Figure~\ref{fig:bestfit_drift}) is a much better (although far from perfect) match to the 0.9--1.4~$\micron$ region as compared to the best-fit BT-Settl model, but  a poorer match to the full near-infrared SED.  For the median parameters, the evolutionary models of \citet{2003A&A...402..701B} predict M = 0.088$^{+0.005}_{-0.008}$~{\msun}, $\tau$ = 0.6$^{+0.9}_{-0.4}$~Gyr and R$_{evol}$ = 0.120$^{+0.017}_{-0.007}$~{\rsun}, which is biased toward higher masses and older ages by the {\logg} constraint; again, the \citet{2001RvMP...73..719B} models yield similar values. Despite the smaller evolutionary model radius, there remains a 2.8$\sigma$ disagreement between it and the scaling factor radius.

\section{Discussion}

Our spectral model fit analysis yields mixed results in elucidating the roles of surface gravity and metallicity in shaping the peculiar spectrum of GJ~660.1B. The models affirm a subsolar metallicity for this source, bracketing the metallicity {determined for} the primary.  However, the BT-Settl models also converge on surface gravities that are appropriate for a very young brown dwarf ($\lesssim$100~Myr), which disagrees with the kinematics of the system and comparison of the spectrum of GJ~660.1B to young template spectra. In addition, neither of the model sets produce a ``best-fit'' that is an accurate representation of the data. The low metallicity, low surface gravity BT-Settl model roughly reproduces the $\lambda$ $>$ 1.4~$\micron$ SED of GJ~660.1B, but fails to match the molecular and atomic features for $\lambda$ $<$ 1.4~$\micron$.  The high metallicity, high surface gravity Drift model has the opposite problem. 

\citet{2009ApJ...706.1114B} reported similar disagreements in their analysis of the d/sdM9 companion
to the metal-poor F9 star of HD~114762 ([Fe/H] = $-$0.7). Their Phoenix/GAIA model fits \citep{2005ESASP.576..565B} to low-resolution near-infrared spectra of HD~114762B converged to very low surface gravities, and retained a high $\chi^2$.  The study concluded that spectral models did not fit the low-resolution spectra of these low-metallicity, low-temperature sources, whereas moderate resolution spectra yielded better fits. \citet{2009ApJ...697..148B}, \citet{2011A&A...529A..44W} and \citet{2014A&A...564A..55M} have found comparable problems for late-M and L dwarf and subdwarf low-resolution spectra and photometry.

What is causing the models to fail?
We conjecture that the interplay of two opacity sources are primarily responsible:  condensate cloud opacity and H$_2$ CIA. Condensates become an important opacity source in the atmospheres of late M and L dwarfs, producing a scattering haze that suppresses the 0.9--1.4~$\micron$ continuum \citep{2001ApJ...556..357A,2001ApJ...556..872A,2003ApJ...586.1320C,2008MNRAS.391.1854H}.  \citet{2003ApJ...592.1186B,2007ApJ...657..494B} and \citet{2006AJ....132.2372G} have conjectured that condensate production is inhibited in metal-poor late M and L subdwarfs, based on their exceptionally blue near-infrared SEDs ($J-K \approx 0$) and persistent TiO optical bands. 
This has been confirmed in theoretical calculations by \citet{2009A&A...506.1367W}, which demonstrated that grain production persists in low metallicity dwarfs but with an overall opacity that declines with metallicity; and by the absolute $J$ magnitudes of subdwarfs, which are brighter than equivalently classified dwarfs \citep{2009A&A...493L..27S,2012ApJ...752...56F}. 
Pressure-sensitive H$_2$ CIA, on the other hand, strengthens in both higher surface gravity and metal-poor dwarfs ($P \propto g/\kappa$), suppressing $K$-band flux and producing bluer SEDs. Both of these trends---reduced condensate opacity and stronger H$_2$ absorption---qualitatively explain the bluer SED tilt and stronger FeH and alkali absorption in the spectrum of GJ~660.1B and other metal-poor dwarfs as compared to field standards (Figure~\ref{fig:classification}).  

However, these opacity sources do not appear to be accurately represented in the models.  Analysis of field M and L dwarf SPL spectra has shown that the Drift models, even with a prescription for metallicity-dependent condensate grain formation, underpredict condensate opacity in dwarfs with {\teff} $>$ 1900~K, resulting in SEDs that are bluer than observed \citep{2011A&A...529A..44W}.  Similarly, \citet{2011ASPC..448...91A} have commented that BT-Settl models, and perhaps ``all the current cloud models'' fail to produce sufficient condensate absorption at the M dwarf/L dwarf transition.
Separately, new calculations by \citet{2012JChPh.136d4319A} and \citet{2010MolPh.108.2265F} indicate current models may be excessively blue in the near-infrared due to incorrect H$_2$ CIA opacity. \citet{2012ApJ...750...74S} estimate that revised opacities translate into a 44\% decrease in $K$-band absorption at 2000~K for cloud-free atmosphere models.

We surmise that both insufficient condensate opacity and excessively strong H$_2$ CIA are present in the Drift and BT-Settl models examined here, which are too blue and therefore biasing fits toward lower surface gravities (BT-Settl) and higher metallicities (Drift). These biases are apparent when we compare our GJ~660.1B spectrum to models for an ``expected'' parameter set ({\teff} = 2600~K, {\logg} = 5.3 and [M/H] = $-$0.65), which exhibit excessively strong molecular bands over 1.0--1.2~$\micron$, excess flux at $H$-band and excess continuum suppression at $K$-band (Figures~\ref{fig:bestfit_btsettled} and~\ref{fig:bestfit_drift}). Increasing condensate opacity in the models would mute the molecular gas features, while both grain scattering and weaker H$_2$ opacities would produce a redder tilt to the SED without the need for lower surface gravities.
GJ~660.1B, HD~114762B, and other 
late M and L dwarf companions to metal-poor stars will be useful benchmarks for testing updated condensate and H$_2$ CIA opacities in future atmosphere models.

Finally, we comment on the most interesting spectral peculiarity of GJ~660.1B, its triangular-shaped $H$-band continuum. As discussed above, this feature is typically associated with low surface gravity sources and attributed to reduced H$_2$ CIA in a low-pressure atmosphere. However, it has also been reported in the spectra of high proper motion late-M dwarfs with no evidence of youth \citep{2010ApJS..190..100K}, and is present in 
the spectra of both GJ~660.1B and HD~114762B \citep{2009ApJ...706.1114B}. These common features point to subsolar metallicity as the likely source. However,
more metal-poor late-M and L subdwarfs do not exhibit this feature (\citealt{2003ApJ...592.1186B,2009ApJ...697..148B,2010ApJS..190..100K,2014ApJ...783..122K}; {Figure~\ref{fig:sdcomp}}). 
An additional opacity source that may help shape the $H$-band continuum of mild subdwarfs is the 1.55--1.60~$\micron$ $E\,^4\Pi-A\,^4\Pi$ FeH absorption band \citep{2003ApJ...582.1066C,2003ApJ...594..651D}. Like its 1.0~$\micron$ and 1.2~$\micron$ counterparts, we expect this feature to be enhanced in the spectra of mild subdwarfs, resulting in a sloped, rather than flat, $H$-band continuum.
Stronger H$_2$O absorption, a consequence of reduced condensate opacity, may also play a role in shaping this feature, while simultaneously skewing index-based near-infrared classifications to later types (Section~3.1).
The presence of this feature, and its influence on the gravity indices, counters the claim of \citet{2013MmSAI..84.1089A} that metal-poor dwarfs cannot be misclassified as low gravity by this technique.  However, we concur with these authors that ``caution should be used when determining'' near-infrared spectral types with indices alone. 
We advocate full-spectum comparisons as a more robust method for characterizing VLM dwarfs with unusual spectral and physical properties.

\acknowledgements
A.\ Burgasser and J.\ Faherty were Visiting Astronomers at the Infrared Telescope Facility, which is operated by the University of Hawaii under Cooperative Agreement no. NNX-08AE38A with the National Aeronautics and Space Administration, Science Mission Directorate, Planetary Astronomy Program.  The authors thank IRTF operators Paul Sears and Brian Cabreira, and instrument specialist John Rayner, for their assistance with the SpeX observations.
We also thank John Johnson for suggesting metallicity analysis of GJ~660.1A, and our anonymous referee for their rapid review.
CA acknowledges funding by the University of California Office of the President through the UC-HBCU program.
The material presented here is based on work supported by the National Aeronautics and Space Administration under Grant No.\ NNX15AI75G.
This research has made use of the SIMBAD database, operated at CDS, Strasbourg, France; 
NASA's Astrophysics Data System Bibliographic Services;
the M, L, T and Y dwarf compendium housed at \url{http://DwarfArchives.org} 
and the Spex Prism Libraries at \url{http://www.browndwarfs.org/spexprism}. 
The authors recognize and acknowledge the very significant cultural role and reverence that the summit of Mauna Kea has always had within the indigenous Hawaiian community.  We are most fortunate and grateful to have the opportunity to conduct observations from this mountain.

%\bibliographystyle{apj}
%\bibliography{biblibrary}{}

\end{document}